\documentclass[twocolumn,amsmath,amssymb,osajnl]{revtex4}
\usepackage{epsfig}
\begin{document}

\title{Mode-Field Radius of Photonic Crystal Fibers Expressed by the $V$--parameter}

\author{Martin Dybendal Nielsen}
\affiliation{Crystal Fibre A/S, Blokken 84, DK-3460 Birker\o d, 
Denmark\\COM, Technical University of Denmark, DK-2800 Kongens Lyngby, Denmark}

\author{Niels Asger Mortensen and Jacob Riis Folkenberg}
\affiliation{Crystal Fibre A/S, Blokken 84, DK-3460
Birker\o d, Denmark}

\author{Anders Bjarklev}
\affiliation{COM, Technical University of Denmark, DK-2800 Kongens Lyngby, Denmark}

\begin{abstract}
We numerically calculate the equivalent mode-field radius of the fundamental mode in a photonic crystal fiber (PCF) and show that this is a function of the V-parameter only and not the relative hole size. This dependency is similar to what is found for graded-index standard fibers and we furthermore show that the relation for the PCF can be excellently approximated with the same general mathematical expression. This is to our knowledge the first semi-analytical description of the mode-field radius of a PCF.
\end{abstract}


\maketitle

Theoretical descriptions of Photonic Crystal Fibers (PCFs) have traditionally been based on numerical methods such as the plane-wave expansion method,\cite{ref1,ref2} methods employing localized functions,\cite{ref3,ref4} or the multipole method.\cite{ref5,ref6} A numerical approach is generally required due to the complex dielectric cross section of the PCF which makes analytical approaches very difficult and results in the fact that no close-form analytical descriptions of either propagation constants or mode fields are available. In this Letter we investigate the relation between the equivalent mode-field radius of the fundamental mode and a recently proposed formulation of the $V$--parameter for a PCF.\cite{ref7} We show that the mode-field radius is a function of the $V$--parameter only and provide an empirical expression describing this relation having the same mathematical form as known from the description of graded-index standard fibers.\cite{ref8}

The PCF analyzed in this work is an all-silica fiber having a triangular arrangement of circular voids with diameter $d$ running along the full length of the fiber. The voids are placed symmetrically around a central defect which acts as the fiber core consisting of a solid silica region i.e. an omitted air hole. The air-hole matrix, which has a lattice constant, $\Lambda$, functions as a cladding region and the fiber structure is invariant in the longitudinal direction. Such a fiber was first proposed in\cite{ref9} and it was shown that it posses unique properties such as endlessly single-mode operation.\cite{ref10} The endlessly single-mode operation is a consequence of the fact that the number of guided modes is finite regardless of wavelength and that the upper limit for this number decreases with the air-filling fraction of the structure i.e. with the value of $d/\Lambda$. Sufficiently small air holes will cause the number of allowed modes to equal two, namely the two degenerate polarization states of the fundamental mode.\cite{ref10} 

In the case of graded-index standard fibers, of which the step-index fiber (SIF) can be considered a special case, the $V$--parameter plays a central role in the description of the number of guided modes, the cut-off criterion [11] as well as the mode-field radius.\cite{ref8} In attempt to obtain an expression for a $V$-parameter adequate for PCFs, approximations based on an equivalent SIF has generally been employed [10,12]. Although this approximation contains the overall correct physics it fails to describe the cut-off properties and has difficulty when it comes to determining an appropriate equivalent core radius. Recently, we proposed a definition of the $V$--parameter for a PCF,\cite{ref7} $V_{\rm PCF}$, rejecting the SIF approximation [See Ref.~\onlinecite{ref7} for a more detailed discussion on Eq. (1)]:
 
\begin{equation}
V_{\rm PCF}(\lambda)=\frac{2\pi}{\lambda}\Lambda \sqrt{n_{\rm co}^2(\lambda)-n_{\rm cl}^2(\lambda)}
\end{equation}

In Eq. (1), $\lambda$ is the free space wavelength and $n_{\rm co}(\lambda)$ and $n_{\rm cl}(\lambda)$ are the effective indices of the fundamental mode and the first eigenmode in the perfect infinite cladding structure, respectively. The latter is often denoted the fundamental space-filling mode and can be interpreted as the refractive index of the cladding material in the absence of the core.\cite{ref10} Using this definition it can be showen that the condition for the  higher-order mode cut-off can be formulated as $V_{\rm PCF} =\pi$.\cite{ref7} Both effective indices in Eq. (1) are strongly wavelength dependent and can not be approximated by constants as in the case of standard fibers. It is due to the strong dispersion of the effective indices that the PCF can be attributed many of its unique properties. In Fig. 1, $V_{\rm PCF}$ calculated using the plane-wave expansion method\cite{ref13} is plotted as function of the normalized frequency, $\Lambda/\lambda$, for values of $d/\Lambda$  ranging from 0.30 to 0.70 in steps of 0.05. The horizontal dashed line in the plot indicates $V_{\rm PCF} = \pi$  and thereby the single-mode boundary. Since the variation of refractive index of silica, $n_{\rm silica}$, over the transparent wavelength region is in the order of a few \%, the effect of material dispersion will only result in a small shift in the effective indices $n_{\rm co}(\lambda)$ and $n_{\rm cl}(\lambda)$. Furthermore, such a perturbation will shift both further reducing the effect on VPCF. In the calculations we therefore used a fixed value of $n_{\rm silica} = 1.444$, thereby preserving the scale invariance of Maxwell's equations.

\begin{figure}[t!]
\begin{center}
\epsfig{file=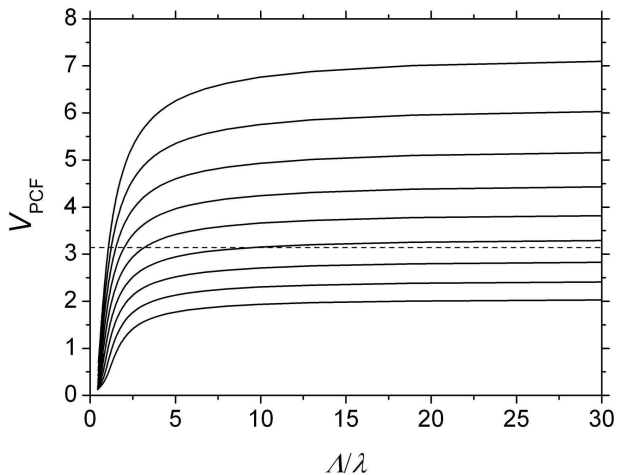, width=0.45\textwidth,clip}
\end{center}
\caption{Calculated $V_{\rm PCF}$ (from Eq. 1) as function of the normalized frequency, $\Lambda/\lambda$, for $d/\Lambda$ equal to 0.30, 0.35, 0.40, 0.45, 0.50, 0.55, 0.60, 0.65, and 0.70 from bottom and up. The dash line indicates the single-mode boundary, $V_{\rm PCF} = \pi$.}
\label{fig1}
\end{figure}

$V_{\rm PCF}$ approaches a constant value, dependent on $d/\Lambda$ for increasing $\Lambda/\lambda$\cite{ref10} and since the number of modes generally increases with the $V$--parameter, the asymptotic behavior of $V_{\rm PCF}$ is consistent with the endlessly single-mode property. In the framework of standard fibers, the weak wavelength dependency of the $V$--parameter will lead to a mode-field radius which is also only weakly dependent on the wavelength.

For a graded-index standard fiber, the index profile can be described by a power law in which case the shape of the index profile is governed by an exponent, $g$. For $g = 1$ the index profile is triangular, for $g = 2$ the profile is parabolic, and in the limit of large values of $g$ the profile approaches that of a SIF. For this type of fiber, the mode field is generally very close to a Gaussian distribution provided that the field does not penetrate too deeply into the cladding region. The agreement between the actual mode and a Gaussian distribution has a weak dependency on $g$ and is a perfect match in the case of a parabolic index profile. The mode-field radius, $w$, is a function of the $V$--parameter and can be fitted using the expression:\cite{ref8}

\begin{equation}
\frac{W}{a}=\frac{{\mathcal A}}{V^{2/(2+g)}}+\frac{\mathcal B}{V^{3/2}}+\frac{\mathcal C}{V^6}
\end{equation}

$\mathcal A$, $\mathcal B$, and $\mathcal C$, are fitting parameters which are dependent on the index profile while a denotes the core radius. In the case of a SIF the first term in Eq. (2) is constant and the values of $\mathcal A$, $\mathcal B$, and $\mathcal C$ are 0.65, 1.619, and 2.879, respectively. The accuracy of the fits provided by Eq. (2) is better than 2\% in the range $1.5 < V < 7$ and the relation is therefore extremely useful when designing and working with SIFs or graded index fibers in general.

In order to investigate a similar relation for the PCF, we first introduce an equivalent mode-field radius of the fundamental mode, $w_{\rm PCF}$, as the mode-field radius of the Gaussian distribution having the same effective area, $A_{\rm eff}$, as the fundamental mode itself, yielding the simple relation $A_{\rm eff} = \pi w^2$. The effective area is calculated as 

\begin{equation}\label{Aeff}
A_{\rm eff}= \Big[\int d{\boldsymbol r}_\perp I({\boldsymbol r}_\perp)\Big]^2\Big[\int d{\boldsymbol r}_\perp I^2({\boldsymbol r}_\perp)\Big]^{-1},
\end{equation}

In Eq. (3), $I({\boldsymbol r}_\perp)$ is the transverse intensity distribution of the mode. Although the intensity distribution of the fundamental mode in a PCF is not rotational symmetric but rather has the 6-fold symmetry of the triangular cladding structure, a Gaussian approximation is in fact very good and has previously been employed in the description of various PCF properties.\cite{ref14,ref15,ref16} By numerical calculation of $A_{\rm eff}$ as function of the normalized wavelength, $\lambda/\Lambda$, the normalized mode-field radius, $w_{\rm PCF}/\Lambda$ can be plotted as function of $V_{\rm PCF}$ as shown in Fig. 2. The normalization with $\Lambda$ is chosen since $\Lambda$ is the natural length scale of the problem in the same sense as the core radius, $a$, is for the graded-index fiber. The plot in Fig. 2 represents data for $d/\Lambda = 0.30$ and $d/\Lambda = 0.70$ indicated by open squares and circles, respectively. The data points for these two calculations overlap in the entire range where the value of $V_{\rm PCF}$ overlap and data for calculations having $d/\Lambda  = 0.35$, 0.40, 0.45, 0.50, 0.55, 0.60, and 0.65 (not shown) fall on the same curve indicating that $w_{\rm PCF}/\Lambda$ is in fact a function of the $V_{\rm PCF}$ only. Also included in Fig. 2 is the corresponding curve for the SIF (dashed line) showing $w_{\rm SIF}/a$ as function of $V_{\rm SIF}$ calculated from Eq. (2). 

\begin{figure}[t!]
\begin{center}
\epsfig{file=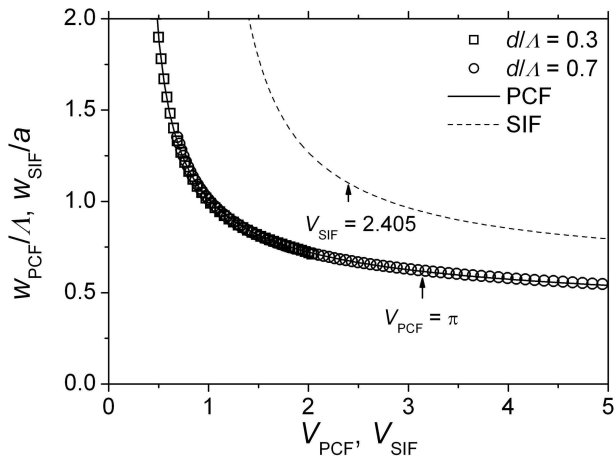, width=0.45\textwidth,clip}
\end{center}
\caption{Normalized mode-field radius for the PCF, $w_{\rm PCF}/\Lambda$, as function of $V_{\rm PCF}$ for $d/\Lambda  = 0.30$, (squares) and $d/\Lambda  = 0.70$ (circles) along with a numerical fit through these points (full line). The corresponding expression for the normalized mode-field radius, $w_{\rm SIF}/a$ for the SIF as function of $V_{\rm SIF}$ is also shown (dashed line). The two arrows indicate the higher-order mode cut off of the SIF and PCF at $V_{\rm SIF} = 2.405$ and $V_{\rm PCF} =\pi$, respectively.}
\label{fig2}
\end{figure}

The functional dependency of $w_{\rm PCF}/\Lambda$ on $V_{\rm PCF}$ is seen to follow the same overall behavior as $w_{\rm SIF}/a$ as function of $V_{\rm SIF}$. For increasing values of $V_{\rm PCF}$, $w_{\rm PCF}/\Lambda$ must saturate at a constant value corresponding to the minimum obtainable mode size whereas it grows dramatically for sufficiently small values of $V_{\rm PCF}$ indicating that the mode is weakly confined to the core and begins to penetrate into the cladding region. This is in good agreement with the fact that small values of $V_{\rm PCF}$ are realized either for small values of $d/\Lambda$ or if the optical wavelength is similar to the structural dimensions, $\lambda\sim\Lambda$.

Due to the similar functional dependency of the mode-field radii on the respective $V$--parameters it is reasonable to use the same fitting function for the PCF as used for graded-index standard fibers. However, if assuming the simplest case of a SIF it is not possible to obtain a good fit and the more general expression, for which $g$ can assume any positive value, was therefore employed. This results in the parameters ${\mathcal A} = 0.7078$, ${\mathcal B} = 0.2997$, and ${\mathcal C} = 0.0037$, respectively in the case for $g = 8$. The fit based on these parameters are plotted in Fig. 2 (solid line) and is seen to be very good for small as well as for large values of $V_{\rm PCF}$. The maximal deviation between values predicted by the fit and the data points is determined to be less than 1\%.

An important difference between the PCF and the SIF can be learned from the curves in Fig. 2 on which the higher-order mode cut-off are indicated at $V_{\rm PCF} = \pi$  and $V_{\rm SIF} = 2.405$, respectively. Whereas the cut-off for the PCF is located at a point where the curve is relatively flat the SIF cut off is located where the curve is much steeper. Also, $V_{\rm SIF}$ depends on the wavelength as  $\lambda^{-1}$ further increasing the wavelength dependency of the mode size in the single-mode region of the SIF. For the PCF, not only the mode-size dependence on $V_{\rm PCF}$ close to cut off is week but also the dependence of $V_{\rm PCF}$ itself on $\lambda$ provided that $\Lambda$ is a few times larger than $\lambda$ or more (See Fig. 1). This results in a mode-field size which is essentially wavelength independent for the PCF compared to the SIF. 

In conclusion, we have shown that the recently proposed expression for the $V$--parameter of a PCF uniquely determines the normalized mode-field radius. Furthermore, we have presented a simple analytical expression describing this relation based on the same general expression used for graded-index standard fibers and thereby further closed the gap between standard fibers and PCF in terms of available tools describing their properties.

\vspace{10mm}
M.D. Nielsen acknowledges financial support by the Danish Academy of Technical Sciences. M.D. Nielsen's e-mail address is mdn@crystal-fibre.com.

\end{document}